\documentclass[10pt,twocolumn]{article}

\usepackage{amssymb,amsmath}
\usepackage[dvips]{graphicx}

\DeclareMathOperator{\Tr}{tr}

\newtheorem{thm}{Theorem}
\newtheorem{prop}[thm]{Proposition}

\newcommand{\vskipline}{\vskip 11pt}

\newcommand{\set}[1]{\lbrace #1 \rbrace}

\newcommand{\intersect}{\cap}

\newcommand{\bra}[1]{\langle #1 |}
\newcommand{\ket}[1]{| #1 \rangle}

\newcommand{\expect}[1]{\langle #1 \rangle}
\newcommand{\norm}[1]{\lVert #1 \rVert}
\newcommand{\tensor}{\otimes}

\newcommand{\RR}{\mathbb{R}}

\newcommand{\PP}{\mathcal{P}}

\begin{document}

\title{Gibbs States and the Consistency of Local Density Matrices}

\author{Yi-Kai Liu\\
Computer Science and Engineering\\
University of California, San Diego\\
\texttt{y9liu@cs.ucsd.edu}}

\date{March 1, 2006}

\maketitle

\begin{abstract}
Suppose we have an $n$-qubit system, and we are given a collection of local density matrices $\rho_1,\ldots,\rho_m$, where each $\rho_i$ describes some subset of the qubits.  We say that $\rho_1,\ldots,\rho_m$ are ``consistent'' if there exists a global state $\sigma$ (on all $n$ qubits) whose reduced density matrices match $\rho_1,\ldots,\rho_m$.  

We prove the following result:  if $\rho_1,\ldots,\rho_m$ are consistent with some state $\sigma \succ 0$, then they are also consistent with a state $\sigma'$ of the form $\sigma' = (1/Z) \exp(M_1+\cdots+M_m)$, where each $M_i$ is a Hermitian matrix acting on the same qubits as $\rho_i$, and $Z$ is a normalizing factor.  (This is known as a Gibbs state.)  Actually, we show a more general result, on the consistency of a set of expectation values $\expect{T_1},\ldots,\expect{T_r}$, where the observables $T_1,\ldots,T_r$ need not commute.  This result was previously proved by Jaynes (1957) in the context of the maximum-entropy principle; here we provide a somewhat different proof, using properties of the partition function.  
\end{abstract}


\section{Introduction}

Many-body systems have an intriguing property:  under the right circumstances, local interactions can conspire to produce long-range or global effects.  This behavior leads to phase transitions in statistical mechanics, and it also appears in combinatorial problems such as 3-SAT.  If we consider quantum systems, the situation is more complicated, due to non-commuting measurements and the possibility of entanglement.  This leads to new kinds of quantum phase transitions \cite{Sachdev}, and new examples such as the Local Hamiltonian problem \cite{Aharonov-Naveh}.  

A basic question in all of these examples is:  if we know local information about various parts of a system, what can we say about the system as a whole?  This paper gives one answer to this question, for quantum systems.  

Suppose we have an $n$-qubit system, and we are given a collection of local density matrices $\rho_1,\ldots,\rho_m$, where each $\rho_i$ describes a subset $C_i \subseteq \set{1,\ldots,n}$ of the qubits.  We say that $\rho_1,\ldots,\rho_m$ are ``consistent'' if there exists a global state $\sigma$ (on all $n$ qubits) whose reduced density matrices match $\rho_1,\ldots,\rho_m$; in other words, for all $i = 1,\ldots,m$, $\Tr_{\set{1,\ldots,n}-C_i}(\sigma) = \rho_i$.  

Clearly, if $\rho_1,\ldots,\rho_m$ are consistent, then whenever two density matrices $\rho_i$ and $\rho_j$ describe overlapping subsets of qubits ($C_i \intersect C_j \neq \emptyset$), they must agree on the intersection $C_i \intersect C_j$; that is, $\Tr_{C_i-(C_i \intersect C_j)}(\rho_i) = \Tr_{C_j-(C_i \intersect C_j)}(\rho_j)$.  This gives a necessary condition for consistency.  

However, the above condition is not sufficient to guarantee consistency.  To see this, consider the following example:  we have three qubits, and we are told that qubits 1 and 2 are in the Bell state $\ket{\Phi^+} = (\ket{00}+\ket{11})/\sqrt{2}$, and qubits 2 and 3 are also in the same state $\ket{\Phi^+}$.  More formally, let $\rho_A = \ket{\Phi^+}\bra{\Phi^+}$, $A = \set{1,2}$, and let $\rho_B = \ket{\Phi^+}\bra{\Phi^+}$, $B = \set{2,3}$.  In this case, $\rho_A$ and $\rho_B$ both agree on qubit 2, since $\Tr_1(\rho_A) = I/2 = \Tr_3(\rho_B)$.  But there is no state $\sigma$ on all three qubits such that $\Tr_3(\sigma) = \rho_A$ and $\Tr_1(\sigma) = \rho_B$; one way to see this is to apply the strong subadditivity inequality, $S(1,2,3) + S(2) \leq S(1,2) + S(2,3)$.  

Thus the consistency of $\rho_1,\ldots,\rho_m$ would seem to be a more subtle question.  We prove the following result:  
\begin{thm}\label{consistent-rhos}
If $\rho_1,\ldots,\rho_m$ are consistent with some state $\sigma \succ 0$, then they are also consistent with a state $\sigma'$ of the form $\sigma' = (1/Z) \exp(M_1+\cdots+M_m)$, where each $M_i$ is a Hermitian matrix acting on the qubits in $C_i$, and $Z = \Tr(\exp(M_1+\cdots+M_m))$.  
\end{thm}
Here, $\sigma \succ 0$ means that $\sigma$ is a positive definite matrix.  The state $\sigma'$ is known as a Gibbs state.  

Essentially, this result says that a Gibbs state $\sigma'$ can simulate an arbitrary state $\sigma \succ 0$, with respect to an observer who can only access subsets $C_1,\ldots,C_m$ of the qubits.  For example, consider a physical system with local interactions, described by a Hamiltonian $H$.  It is easy to see that the ground state of $H$ can be approximated by $\eta = (1/Z) \exp(-\beta H)$, for $\beta$ large; and since $H$ is a sum of local terms, $\eta$ is a Gibbs state.  Our result extends this simple observation to a much more general setting.  

Actually, we prove the following more general result:  Consider a finite quantum system, and let $T_1,\ldots,T_r$ be observables (Hermitian matrices).  Without loss of generality, assume that the collection of matrices $I,T_1,\ldots,T_r$ is linearly independent (over $\RR$).  We say that a state $\rho$ has expectation values $t_1,\ldots,t_r$ if $\Tr(T_i \rho) = t_i$ for all $i = 1,\ldots,r$.  
\begin{thm}\label{consistent-ts}
If there exists some state $\rho \succ 0$ which has expectation values $t_1,\ldots,t_r$, then there exists a state $\rho'$ which has the same expectation values $t_1,\ldots,t_r$, and is of the form $\rho' = (1/Z) \exp(\theta_1 T_1 + \cdots + \theta_r T_r)$, where $\theta_1,\ldots,\theta_r \in \RR$.  
\end{thm}
This statement holds even when the observables $T_1,\ldots,T_r$ do not commute.  

This result was previously proved by Jaynes, as part of the maximum entropy principle in statistical mechanics \cite{Jaynes-2,Jaynes-Brandeis}.  Jaynes showed that the Gibbs state $\rho'$ is the state which maximizes the entropy $S(\rho) = -\Tr(\rho \log \rho)$ subject to the constraints $\expect{T_i} = t_i$; implicitly, he also showed that the Gibbs state $\rho'$ is always feasible, in the sense that it can produce the same expectation values $\expect{T_i}$ as an arbitrary state $\rho \succ 0$.  

However, Jaynes' motivation was somewhat different from ours.  Jaynes was interested in statistical mechanics, which deals with large systems with many degrees of freedom and only a few constraints.  Feasibility is not usually a concern in such cases, while the maximum-entropy property is crucial in making plausible inferences about the ``true'' state of the system.  

In this paper, we focus on finite quantum systems, with many non-commuting constraints; we are interested in the relationship between local constraints and the global state of the system.  For us, feasibility of the Gibbs state is important, since it is possible for the system to become overdetermined.  Statistical inference is less important, because the systems we study are small enough that their state can be completely determined (at least in principle).  Rather than viewing this as an inference problem, we can speak directly about what states are allowed under a given set of constraints.  

Finally, we prove our result using a technique which is different from Jaynes.  Jaynes used the Lagrange dual of the entropy-maximization problem, while we use some analytic properties of the partition function.  Our analysis bears some resemblance to classical results on exponential families in statistics \cite{Brown}---although the technical details are quite different.  Our proof also contains some geometric intuition which may be of interest.


\section{Proofs of our results}

First, we will review some useful facts about the partition function for a Gibbs state.  Then we will prove theorem \ref{consistent-ts}, and obtain theorem \ref{consistent-rhos} as a special case.  

\subsection{The partition function}

Recall the situation described in theorem \ref{consistent-ts}:  we have a finite quantum system, and observables $T_1,\ldots,T_r$, such that $I,T_1,\ldots,T_r$ are linearly independent (over $\RR$).  We are interested in states of the form 
\[
\rho(\theta) = \exp(\theta_1 T_1 + \cdots + \theta_r T_r) / Z(\theta), 
\quad \theta \in \RR^r, 
\]
where $Z(\theta) = \Tr(\exp(\theta_1 T_1 + \cdots + \theta_r T_r))$.  $Z(\theta)$ is called the partition function, and we also define the log partition function $\psi(\theta) = \log Z(\theta)$.  

Note that, in the above definition, we can translate each observable $T_i$ by a multiple of the identity, without changing the state $\rho(\theta)$.  More precisely, if we define new observables $P_i = T_i + \lambda_i I$, with $\lambda_i \in \RR$, we have that:  
\[
\frac{\exp(\theta_1 P_1 + \cdots + \theta_r P_r)}
     {\Tr(\exp(\theta_1 P_1 + \cdots + \theta_r P_r))}
 = \frac{\exp(\theta_1 T_1 + \cdots + \theta_r T_r)}
        {\Tr(\exp(\theta_1 T_1 + \cdots + \theta_r T_r))}.  
\]
Using subscripts $T$ and $P$ to denote the two sets of observables, we arrive at the same state, $\rho_P(\theta) = \rho_T(\theta)$, although the partition functions are different, $Z_P(\theta) \neq Z_T(\theta)$.  

The log partition function $\psi$ has some nice analytic properties:  it is convex, and its derivatives encode the expectation values of the observables $T_i$.  We briefly sketch these results, which can be found in quantum statistical mechanics \cite{Jaynes-Brandeis}, as well as quantum information geometry \cite{Ingarden}.  

\begin{prop}\label{convexity}
$\psi$ is convex on $\RR^r$.  
\end{prop}
Proof sketch:  This follows from some facts in matrix analysis \cite{Bhatia}.  First, the Golden-Thompson inequality:  If $A$ and $B$ are Hermitian matrices, then 
\[
\Tr(\exp(A+B)) \leq \Tr(\exp(A)\exp(B)).  
\]
Next, a matrix version of H\"older's inequality:  For any matrix $A$, define the Frobenius or Hilbert-Schmidt norm to be $\norm{A}_2 = (\Tr(A^\dagger A))^{1/2}$.  Also, let $|A|$ denote the unique positive semidefinite square root of $A^\dagger A$.  Then we have that, for all square matrices $A$ and $B$, 
\[
\norm{AB}_2 \leq \norm{|A|^p}_2^{1/p} \norm{|B|^q}_2^{1/q}, 
\]
for $\tfrac{1}{p} + \tfrac{1}{q} = 1$, $p > 1$.  $\square$

\begin{prop}\label{derivatives}
$\psi$ is differentiable on $\RR^r$, and 
\[
\frac{\partial\psi}{\partial\theta_i} = \Tr(T_i \rho(\theta)) = \expect{T_i}.  
\]
\end{prop}
Proof sketch:  Use ``parameter differentiation'' \cite{Wilcox}:  If $H$ is a Hermitian matrix which depends on a parameter $\lambda$, and $\partial H / \partial \lambda$ and $\partial^2 H / \partial \lambda^2$ exist and are continuous, then $\partial(\exp(H)) / \partial \lambda$ exists and is equal to 
\[
\frac{\partial}{\partial\lambda} \exp(H)
 = \int_0^1 \exp((1-u)H) \frac{\partial H}{\partial\lambda} \exp(uH) du.  \quad \square
\]


\subsection{Proof of theorem 2}

Proof:  We are given expectation values $t_1,\ldots,t_r$, and we want to find a state 
\[
\rho'(\theta) = \exp(\theta_1 T_1 + \cdots + \theta_r T_r) / Z'(\theta) 
\]
that has these expectation values.  (Here, $Z'(\theta)$ is the partition function, and $\psi'(\theta) = \log Z'(\theta)$ is the log partition function.)  By translating the observables $T_i$, we can assume that $t_i = 0$, for all $i = 1,\ldots,r$.  We can now restate the problem in terms of the log partition function:  we are looking for some $\theta \in \RR^r$ such that $\nabla \psi'(\theta) = 0$.  

We know there exists a state $\rho \succ 0$ which has the desired expectation values $t_1,\ldots,t_r$.  Now choose some observables $U_1,\ldots,U_s$, such that the set $\set{I,T_1,\ldots,T_r,U_1,\ldots,U_s}$ is complete and linearly independent (in other words, any $2^n$-dimensional Hermitian matrix can be written uniquely as a real linear combination of the matrices in this set).  Let $u_1,\ldots,u_s$ be the expectation values of $\rho$ for the observables $U_1,\ldots,U_s$; that is, $u_i = \Tr(U_i \rho)$.  By translating the $U_i$, we can assume that $u_i = 0$, for all $i = 1,\ldots,s$.  

We will consider states of the form 
\[
\begin{split}
\rho(\theta,\phi)
 = \exp\bigl( & \theta_1 T_1 + \cdots + \theta_r T_r + \\
              & \phi_1 U_1 + \cdots + \phi_s U_s \bigr) / Z(\theta,\phi).  
\end{split}
\]
(Here, $Z(\theta,\phi)$ is the partition function, and $\psi(\theta,\phi) = \log Z(\theta,\phi)$ is the log partition function.)  Completeness of the $T_i$ and the $U_i$ implies that we can write $\rho$ in the form $\rho = \rho(\theta,\phi)$ for some $(\theta,\phi) \in \RR^{r+s}$.  This implies that $\nabla \psi(\theta,\phi) = 0$ for some $(\theta,\phi) \in \RR^{r+s}$.  

Furthermore, we claim that there is a unique point $(\theta,\phi)$ such that $\rho(\theta,\phi)$ has the expectation values $t_i$ and $u_i$.  This is because the expectation values $t_i$ and $u_i$ uniquely determine the state $\rho$, and setting $\rho = \rho(\theta,\phi)$ uniquely determines the values of $\theta$ and $\phi$.  This in turn follows from the completeness and linear independence of the $T_i$ and the $U_i$.  So we conclude that $\nabla \psi(\theta,\phi) = 0$ at exactly one point $(\theta,\phi)$.  

\begin{figure}
\begin{center}
\includegraphics{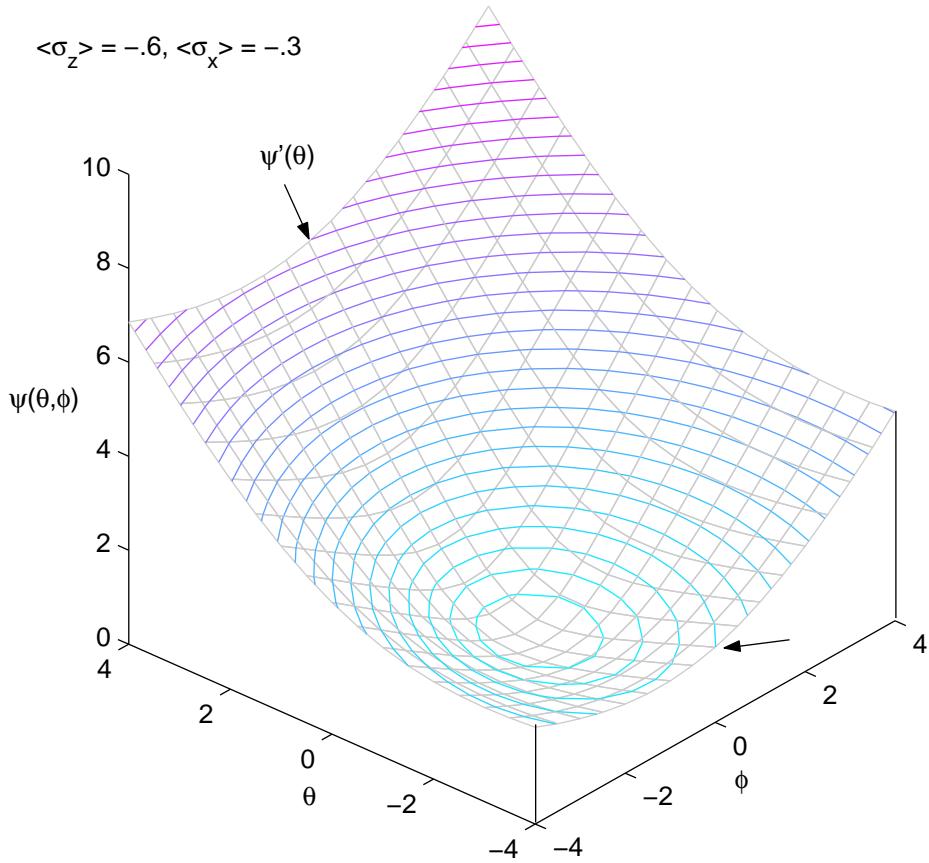}
\caption{A single-qubit example.  We want to find a Gibbs state $\rho'$ that satisfies $\expect{\sigma_z} = -0.6$; we have one observable $T = \sigma_z + 0.6$.  We know there exists some state $\rho \succ 0$ that satisfies $\expect{\sigma_z} = -0.6$; in this case, $\rho$ also satisfies $\expect{\sigma_x} = -0.3$, and we let $U = \sigma_x + 0.3$ play the role of the ``extra'' observables.  As the graph shows, $\nabla \psi(\theta,\phi)$ vanishes at exactly one point; $\psi'(\theta) = \psi(\theta,0)$; and $\nabla \psi'(\theta)$ vanishes for some $\theta$.}
\label{fig-geometry}
\end{center}
\end{figure}

To complete the proof, we will carry out the following plan:  we will show that $\psi(\theta,\phi) \rightarrow \infty$ as $\norm{\theta,\phi} \rightarrow \infty$, where $\norm{\theta,\phi}$ denotes the norm of the vector $(\theta,\phi)$.  This implies that $\psi'(\theta) \rightarrow \infty$ as $\norm{\theta} \rightarrow \infty$; and hence $\nabla \psi'(\theta) = 0$ for some $\theta \in \RR^r$.  (See figure \ref{fig-geometry} for a simple example that shows the geometric intuition for the proof.)  

Let $(\theta_0,\phi_0)$ be the unique point where $\nabla \psi$ vanishes.  We claim that $(\theta_0,\phi_0)$ is the unique global minimum of $\psi$.  [Since $\psi$ is convex (proposition \ref{convexity}), it follows that $\psi$ is bounded below, and $(\theta_0,\phi_0)$ is a global minimum.  Also, $\psi$ is differentiable everywhere on the domain $\RR^{r+s}$, which has no boundaries (proposition \ref{derivatives}); so any extremum $(\theta,\phi)$ must satisfy $\nabla \psi(\theta,\phi) = 0$.  But this happens only at $(\theta_0,\phi_0)$, and so $(\theta_0,\phi_0)$ is the unique global minimum.]  

Let $S$ be the set of all unit vectors in $\RR^{r+s}$.  Define the function $f(\nu,z) = \psi((\theta_0,\phi_0) + z\nu)$, for $\nu \in S$, and $z \in \RR$.  Say we fix $z = 1$.  We claim that there exists some $b > 0$ such that, for all $\nu$, $f(\nu,1) \geq \psi(\theta_0,\phi_0) + b$.  [Since $(\theta_0,\phi_0)$ is the unique global minimum, we have that $f(\nu,1) > \psi(\theta_0,\phi_0)$, for all $\nu$.  Moreover, $f(\nu,1)$ is a continuous function of $\nu$, and $S$ is a compact set, hence its image $f(S,1)$ is compact.  Hence $f(\nu,1)$ must be bounded away from $\psi(\theta_0,\phi_0)$, for all $\nu$.]  

Next we claim that, for all $\nu$, and for all $z \geq 1$, $(\partial f / \partial z)(\nu,z) \geq b$.  [Fix any $\nu$.  $f(\nu,z)$ is a differentiable function of $z$, so by the mean value theorem, there exists some $z \in (0,1)$ such that $(\partial f / \partial z)(\nu,z) = f(\nu,1) - f(\nu,0) \geq b$.  In addition, since $\psi$ is convex, $(\partial f / \partial z)(\nu,z)$ is nondecreasing in $z$.  This proves the claim.]  

Now, say we are given some $(\theta,\phi)$, and assume that $\norm{(\theta,\phi)-(\theta_0,\phi_0)} \geq 1$.  We can write $(\theta,\phi)$ in the form 
\[
(\theta,\phi) = (\theta_0,\phi_0) + \norm{(\theta,\phi)-(\theta_0,\phi_0)} \nu, 
\]
for some unit vector $\nu \in S$.  Then we have:  
\[
\begin{split}
\psi(\theta,\phi)
 &= f(\nu,\norm{(\theta,\phi)-(\theta_0,\phi_0)})\\
 &= f(\nu,1) + \int_1^{\norm{(\theta,\phi)-(\theta_0,\phi_0)}} 
               (\partial f / \partial z)(\nu,z) dz\\
 &\geq \psi(\theta_0,\phi_0) + b + b (\norm{(\theta,\phi)-(\theta_0,\phi_0)} - 1)\\
 &= \psi(\theta_0,\phi_0) + b \norm{(\theta,\phi)-(\theta_0,\phi_0)}.  
\end{split}
\]
From this, we conclude that $\psi(\theta,\phi) \rightarrow \infty$ as $\norm{\theta,\phi} \rightarrow \infty$.  

Notice that the partition functions for $\rho'(\theta)$ and $\rho(\theta,\phi)$ are related:  
\[
\psi'(\theta) = \psi(\theta,0).  
\]
Hence, $\psi'(\theta) \rightarrow \infty$ as $\norm{\theta} \rightarrow \infty$.  

We will use the following fact:  if $f:\; \RR^n \rightarrow \RR$ is continuous, and $f(x) \rightarrow \infty$ as $\norm{x} \rightarrow \infty$, then $f$ is bounded below, and $f$ attains its minimum at some point $x_* \in \RR^n$.  [To see this, let $S = \set{x \in \RR^n \;|\; f(x) \leq \alpha}$, choosing $\alpha$ large enough that $S \neq \emptyset$.  Note that $S$ is bounded; otherwise, there would exist a sequence $\set{x_i}$ such that $\norm{x_i} \rightarrow \infty$ and $f(x_i) \leq \alpha$, a contradiction.  Also, note that $S$ is closed; this is because the interval $(-\infty,\alpha]$ is closed, and $f$ is continuous.  So we have that $S$ is compact.  This implies that $f(S)$ is compact.  Hence $f(S)$ is closed and bounded; also note that $f(S) \neq \emptyset$.  This implies that $f$ is bounded below, and attains its minimum.]  

From this, we conclude that $\psi'$ attains its minimum at some point $\theta_* \in \RR^r$.  $\RR^r$ has no boundaries, and $\psi'$ is differentiable everywhere on $\RR^r$, so it follows that $\nabla \psi'(\theta_*) = 0$.  This completes the proof.  $\square$


\subsection{Proof of theorem \ref{consistent-rhos}}

Proof:  We will obtain theorem \ref{consistent-rhos} as a special case of theorem \ref{consistent-ts}.  The basic idea is that specifying the local density matrices $\rho_1,\ldots,\rho_m$ is equivalent to specifying the expectation values of all Pauli matrices on the subsets $C_1,\ldots,C_m$.  

Let $X$, $Y$ and $Z$ denote the Pauli matrices for a single qubit, and define $\PP = \set{I,X,Y,Z}$.  We can construct $n$-qubit Pauli matrices by taking tensor products $P = P_1 \tensor \cdots \tensor P_n \in \PP^{\tensor n}$.  Any $2^n$-dimensional Hermitian matrix can be written as a real linear combination of $n$-qubit Pauli matrices.  Furthermore, the $n$-qubit Pauli matrices are orthogonal with respect to the Hilbert-Schmidt inner product:  $\Tr(P^\dagger Q) = 2^n$ if $P = Q$, and 0 otherwise.  

We make the following claim:  Let $\sigma$ be a density matrix on $n$ qubits, and let $\rho$ be a density matrix on a subset of the qubits $C \subseteq \set{1,\ldots,n}$, with $|C| = k$.  We claim that $\Tr_{\set{1,\ldots,n}-C}(\sigma) = \rho$, if and only if, for all Pauli matrices $P$ on the subset $C$, $\Tr((P \tensor I) \sigma) = \Tr(P \rho)$.  (Notation:  we write $n$-qubit Pauli matrices in the form $P \tensor Q$, where $P$ acts on the subset $C$, and $Q$ acts on the rest of the qubits.)  

The ($\Rightarrow$) direction is obvious, but we need to show ($\Leftarrow$).  We write $\sigma$ and $\rho$ as linear combinations of Pauli matrices, with real coefficients $\beta_{(P \tensor Q)}$ and $\alpha_P$:  
\begin{align*}
\sigma &= \sum_{(P \tensor Q) \in \PP^{\tensor n}} \beta_{(P \tensor Q)} P \tensor Q \\
\rho &= \sum_{P \in \PP^{\tensor k}} \alpha_P P.  
\end{align*}
We know that, for all Pauli matrices $P$ on the subset $C$, $\Tr((P \tensor I) \sigma) = 2^n \beta_{(P \tensor I)} = \Tr(P \rho) = 2^k \alpha_P$.  But this implies:  
\[
\begin{split}
\Tr_{\set{1,\ldots,n}-C}(\sigma)
 &= \sum_{P \in \PP^{\tensor k}} 2^{n-k} \beta_{(P \tensor I)} P \\
 &= \sum_{P \in \PP^{\tensor k}} \alpha_P P = \rho, 
\end{split}
\]
which proves the claim.  

Thus, theorem \ref{consistent-rhos} is a special case of theorem \ref{consistent-ts}, where the observables $T_1,\ldots,T_r$ consist of all the Pauli matrices on the subsets $C_1,\ldots,C_m$.  $\square$


\vskipline

\noindent
\textit{Acknowledgements:}  I am grateful to Dorit Aharonov, Chris Fuchs and David Meyer for helpful discussions about this work.  Funded by an ARO/NSA Quantum Computing Graduate Research Fellowship.



\end{document}